\documentclass[preprint,showpacs,preprintnumbers,amsmath,amssymb,prl]{revtex4}

\usepackage{graphicx}
\usepackage{bm}

\begin{document}

\preprint{}

\title{Revealing effective classifiers through network comparison}

\author{Lazaros K. Gallos}
\author{Nina H. Fefferman}
\affiliation{
Department of Ecology, Evolution, \& Natural Resources, Rutgers University, New Brunswick 
NJ 08901, USA, and DIMACS, Rutgers University, Piscataway NJ 08854, USA}

\date{\today}

\begin{abstract}
The ability to compare complex systems can provide new insight into the fundamental 
nature of the processes captured in ways that are otherwise inaccessible to observation. 
Here, we introduce the $n$-tangle method to directly compare two networks for structural 
similarity, based on the distribution of edge density in network subgraphs. We 
demonstrate that this method can efficiently introduce comparative analysis into network 
science and opens the road for many new applications. For example, we show how the 
construction of a phylogenetic tree across animal taxa according to their social 
structure can reveal commonalities in the behavioral ecology of the populations, or how 
students create similar networks according to the University size. Our method can be 
expanded to study a multitude of additional properties, such as network classification, 
changes during time evolution, convergence of growth models, and detection of structural 
changes during damage.
\end{abstract}

\pacs{89.75.Fb, 89.75.Da, 87.23.Ge}

\maketitle

Advances in quantitative methods for network analysis have allowed researchers across 
fields to quantify and characterize patterns of interaction among individuals, with 
applications in a startling diversity of fields\cite{1}. As in the progression of many 
quantitative tools, while initial efforts to use network analysis were mainly 
descriptive\cite{2}, research then advanced to focus on using them as predictive tools, 
isolating particular characteristics that can provide insight into the system of 
interest\cite{3,4}. However, the richest and most interesting level of investigation from 
new metrics frequently arises when they are ultimately used to make comparisons across 
systems, discovering which characteristics are shared and which are not. The ability to 
compare systems has always been a strong driving force in science\cite{5}. Even the most 
straightforward new tests, such as the discovery of Gram staining for the classification 
of bacterial walls, can lead to breakthroughs that influence generations of research (in 
this case, becoming the cornerstone for progress in drug discovery and antibiotic 
therapies\cite{6}).

Currently, there is not a rigorous definition of network similarity.
This allows similarity to be as broadly interpreted as just one single quantity
averaged over the entire system - e.g. networks with the same average degree -
or it can be extremely restrictive, e.g. node-to-node correspondence in identical networks.
Obviously, no one property can fully characterize a network: for instance, networks can be
structurally very different if they have the same degree distribution but different
clustering coefficient. Even if the clustering coefficient is the same, it is possible
that the networks will have different modularity, etc. It is not known how many and 
which properties should be combined to construct a weighted index of similarity.
Therefore, current research has been directed to alternative methods. Motif comparison \cite{17}
or graphlet comparison \cite{18}, for example, is based on the idea that if we continuously isolate
parts of the network and find the same patterns to occur in the same frequency in two networks,
these networks will have a higher probability of being `similar' to each other. 
However, there are many practical constraints that
render these techniques incapable of handling larger networks or larger motifs \cite{Baskerville}. The most recent advance in the field \cite{Onnela} 
introduced a novel concept in which the network is broken down in communities at different scales and the
comparison is based on network modularity properties.
The question of similarity
under this method becomes `how similar the modular structure of the networks is'.

Here we introduce a measure to detect similarity based on direct topological properties:
Topological Analysis of Network subGraph Link/Edge (tangle) Density. 
Many of these properties can be captured by the distance from a tree structure at different
length scales. The method combines the insight of motifs, simplified for efficiency, and
focuses on microscopic structure compared to the mesoscopic approach of modularity comparison in Onnela et al \cite{Onnela}.
Where the advantage of the motif method is that it takes into account the local configuration of the links,
if we relax the motif requirement for exactly matching patterns we can use the links density as our metric.
The basic foundation of our method is to calculate how the density of links behaves at different scales across
the network. This choice represents many advantages since it incorporates information from many structural properties,
e.g. on the degree (naturally through $\rho=\langle k\rangle/n$), the clustering coefficient ($n$-tangle=3 in our method),
the number of loops (every additional link in a tree structure increases the number of loops), etc.
Additionally the calculation of the similarity index is straightforward and is bounded between 0 and 1.

It is possible to use other properties instead of density, such as the local degree distribution or clustering coefficient,
but the crucial step is the sampling of the connected subgraphs. For example, a specific partition of the network, such as one optimizing
modularity \cite{Ahn} does not contain enough information about the network structure. Similarly, network-wide metrics cannot capture the local details, the
natural inhomogeneity, or possible scale-dependent differences in structure.

Notice that our definition of similarity is `similar local structures', or equivalently that the extended neighborhood
of a node looks similar with the extended neighborhood of another node. In this approach, similarity indicates the number
of loops that exist in continuously expanding scales. Therefore, all tree structures will be deemed equivalent by our method,
even if they are different structures, e.g. a scale-free tree vs an ER tree. In other words, our question for similarity becomes
`how far away is a given structure from a tree' or equivalently `how close is it to a complete subgraph'.
This question is easily calculable and takes into account possible local deviations of the local structure from the global topology.

The crux of this method is to capture how many affiliations we expect to find 
when we isolate any given size of connected sub-group. The concept is the following: 
Consider a connected group of 10 students, which is randomly selected from a class of 100 
students. If you are in this group, how many direct friends do you expect to find in this 
sample, or in other words what is the average edge density in the group? We define this 
to be the 10-tangle density (or $n$-tangle, for any $n$). If we construct the histogram 
of densities from different samples then we can compare these distributions in two 
different networks, and we can know the extent of association in a group of a given size 
independently of the pattern formed in each subgroup. In this way, our method bypasses 
the need to determine direct node-to-node correspondence\cite{15}, while still capturing 
node-level properties of the network for comparison. 

\begin{figure}
\centerline{\resizebox{11cm}{!} { \includegraphics{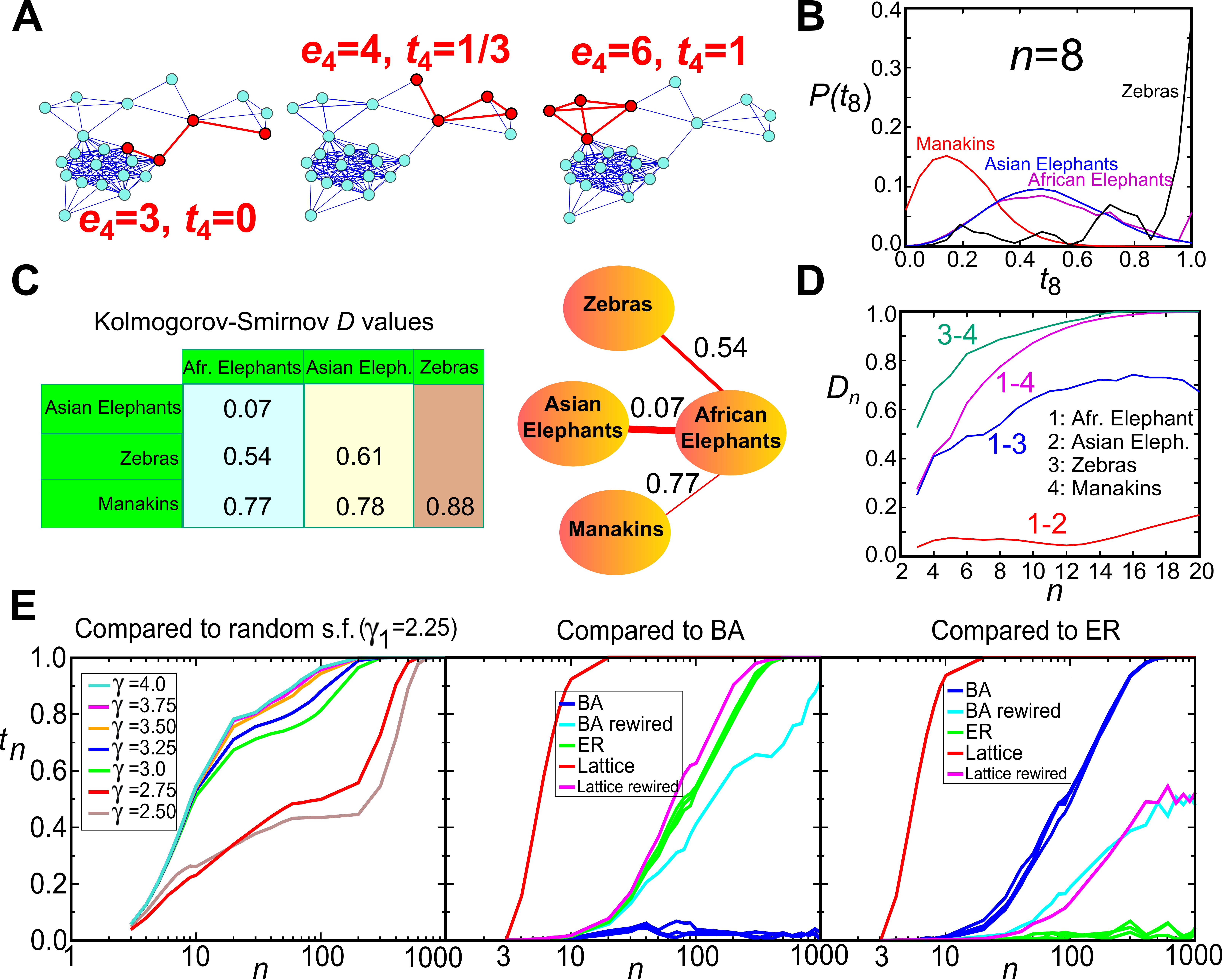}}}
 \caption {
The $n$-tangle method. (A) We randomly sample connected induced subgraphs
of $n$ nodes and calculate their normalized link density $t_n$.
(B) We construct the $n$-tangle histogram $P(t_n)$ for a 
given value of $n$ (the example shows the 8-tangle distribution for 4 animal social networks).
(C) We calculate the distance between any two distributions I and J through,
e.g., a Kolmogorov-Smirnov statistic $D_n(I-J)$. These $D$ values are used as 
the distance between the original networks and can be mapped to
a minimum spanning tree (shown here 
for the four networks), a hierarchical tree, or a threshold-based network.
(D) Variation of the distance between these 4 networks as a function of the subgraph 
size $n$.
(E) The distance of a random scale-free network with a degree exponent $\gamma$ from a network
with $\gamma_1=2.25$ increases monotonically as we increase the value of $\gamma$ (left). Similarly,
the $n$-tangle distance among random Barabasi-Albert networks (center) or random Erdos-Renyi networks
(right) is close to zero, while distances with other model networks are significantly higher.
}
\label{fig1}\end{figure}

Formally, we define the $n$-tangle method in the following way. In a graph $G(V,E)$ 
comprising a set $V$ of nodes and a set $E$ of edges we isolate all possible connected 
induced sub-graphs $G^{\rm in}(V_n,E_n)$. The condition for these sub-graphs is that they should 
include exactly $n$ nodes ($|V_n|=n$) and the subset $E_n$ of $E$ should include all 
$e_n$ links among those $n$ nodes in $G$. For each subgraph we define the $n$-tangle density, 
$t_n$, as the normalized edge density of this subgraph, i.e. the fraction of existing 
over all possible links, after we remove the $n-1$ links that are needed for 
connectivity:
\begin{equation}
t_n = \frac{e_n-(n-1)}{n(n-1)/2-(n-1)} = \frac{2 ( e_n - n + 1)}{n^2-3n+2} \,.
\end{equation}
It is important that the size of the $n$-tangle remains much smaller than the network
size $N$, $n<<N$, so that the sampled subgraphs are statistically independent from each other.
To include the considerably inhomogeneous character of the local structure 
in networks, we consider the $n$-tangle distribution $P(t_n)$ of all $G^{\rm in}$ (Figs~1A, 1B). This 
distribution represents the `signature' of a network at a given subgraph size $n$. We 
repeat this process for all different subgraph sizes $n$, resulting to potentially 
different signatures as we vary $n$. We can then compare the degree of similarity of two 
networks A and B at a given scale by a simple Kolmogorov-Smirnov statistic\cite{16}, 
$D_n(A-B)=\sup|F_A(t_n)-F_B(t_n)|$, where $F_A(t_n)$ is the corresponding $n$-tangle 
cumulative probability in network A and $\sup$ denotes the supremum value (Fig.~1C). Other metrics
of distribution distance can also be used with similar results, (see Supplemental Material).
Since the full comparison involves all subgraph sizes, this method can reveal how two networks can be 
similar at a local scale, while at a larger scale they may exhibit different structures, 
allowing both global network comparison and local analysis of the scale at which 
similarity may be greatest (Fig.~1D).

Our approach avoids the inherent constraints of motif\cite{17} or graphlet\cite{18} based 
methods\cite{19}, by ignoring the costly calculation of the specific pattern created by 
the group and instead placing emphasis on the density of the group, i.e. a single number. 
Therefore, the exponential increase in the number of patterns as a function of group 
size, which limits those techniques to very small-size patterns, does not influence the 
applicability of our method to larger sub-graphs. Now, we only need to keep the number of 
links for each configuration, which makes the calculation and storage very fast. Even 
though the computational complexity of the n-tangle method does increase with the 
subgraph size, the connectedness of bigger social groups can be probed at practically any 
size n, through a fast sampling method. We used a simple Monte-Carlo method to sample a 
large number of configurations, where we repeatedly selected random subgraphs and 
calculated the links within the subgraph.

Our method is designed to quantify local edge densities which combine a lot of 
structural information, and as such it can successfully detect changes in standard network 
properties. In Fig.~1E we compare a series of random scale-free networks created by the
configuration model with a similar network with degree exponent $\gamma_1=2.25$. The
networks become more distant as the exponent of these networks increases, demonstrating that
the method can separate similar structures with different parameters. Similarly, we compare
a number of networks to a sample Barabasi-Albert (BA) network. Different realizations of
BA networks are found to be at almost zero distance with each other, but a randomly rewired BA network
has a different structure. Similarly, lattices and ER networks are also far from the BA network.
Analogous results are found when we compare these model networks with an ER network.

\begin{figure}
\centerline{\resizebox{14cm}{!} { \includegraphics{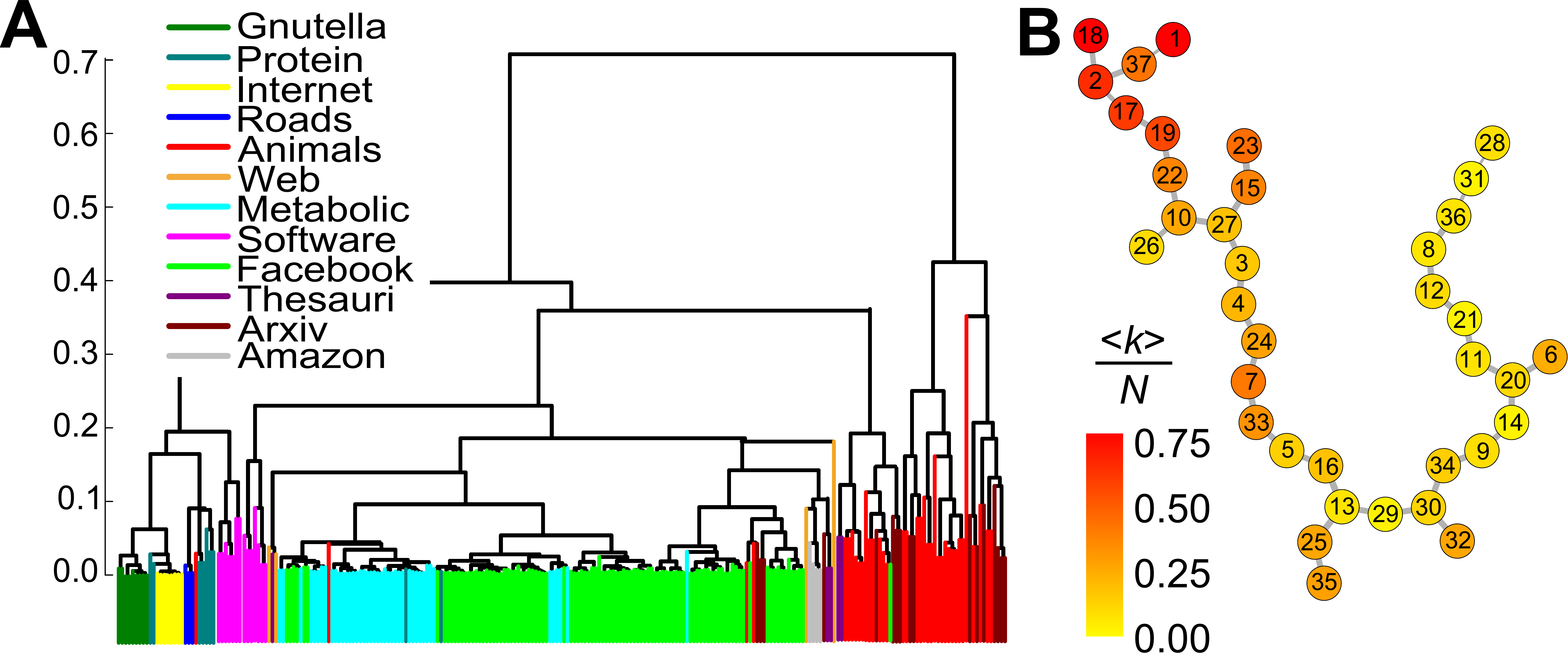}}}
 \caption { 
(A) Hierarchical tree of 236 networks from different fields, based on the $n$-tangle
distance (here $n$=5). 
We used the UPGMA (Unweighted Pair-Group Method 
using Arithmetic Averages) hierarchical clustering method\cite{30}. Colors represent
networks in the same family, as indicated in the index.
(B) The Minimum Spanning Tree for animal networks, based on the $n$-tangle distance
($n=12$).  The species color 
corresponds to varying levels of normalized degree $\langle k \rangle/N$
and separates nicely the species. The node numbers correspond to the species in table I of the SM.
}
\label{fig2}
\end{figure}

We demonstrate the $n$-tangle method first by comparing 236 network
structures of different origins (described in the Supplemental Material). The hierarchical tree in Fig.~2A indicates that networks
from the same family tend to cluster with each other. For example, friendship networks
in facebook are in general closer to each other than with e.g. animal social networks, which also tend to be detected as similar.
We consider this natural separation as a simple verification test for the method.

A more interesting problem is to detect network similarities in systems from within the same family.
For example, we can construct a phylogeny of animal species 
based on their social structure\cite{20}. In this way, we explore whether species with 
similar descriptive characterizations in behavioral ecology do, in fact, exhibit similar 
social structures\cite{21}. We analyze empirically determined contact affiliation 
networks of 33 animal species (described in the Supplemental Material).
In molecular biology, phylogenetic trees can be constructed from 
evolutionary distance\cite{22} (pair wise distances between sequences). Here, rather than 
using species genetic data, our input data are
the pair-wise distances of the $n$-tangle method.

We are therefore able, using our analysis, to
determine whether or not a meaningful cluster results from a choice of a particular facet of the system.
In this example, we find that the normalized 
average degree, i.e. $\langle k \rangle /N$, is able to generate clear clustering by 
$n$-tangle analysis.
This result of our method can provide the first insights 
into whether qualitatively similar social classifications in fact yield similar 
population-level networks of interaction across species (for example, do all dominance 
hierarchies yield similar social structures for the entire population?). This is a 
critical next step in understanding animal social systems. 

\begin{figure}
\centerline{\resizebox{14cm}{!} { \includegraphics{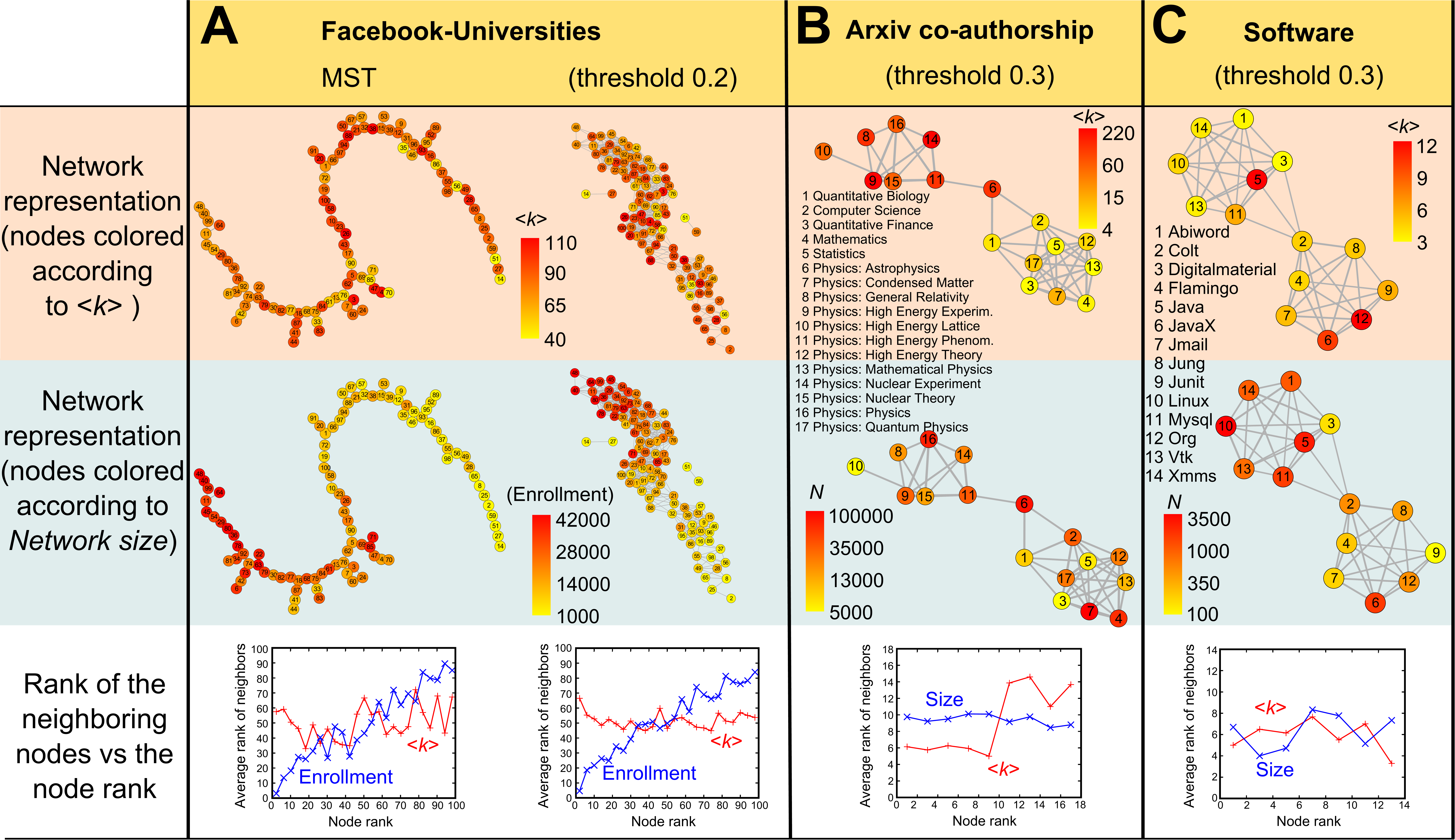}}}
\caption {
Comparison of static networks. (A) Minimum Spanning Tree and threshold-based network 
representation of similarities in the networks of facebook friendship in 100 US 
Universities. The color of the University-nodes corresponds to either the average degree 
or the University size, in terms of enrollment size. The enrollment size is the key 
property for clustering. The plot at the bottom row compares the rank of a University to 
its neighbors rank. The enrollment size has a very hierarchical structure where ranks of 
the same order connect to each other, in contrast to average degree ranking where a 
node’s rank cannot predict the rank of its neighbors. (B) The similarity network of 
scientific fields, based on co-authorship, exhibits the opposite trend. The average 
degree is a nice indicator for clustering, while the network size is not. This result is 
supported by the plot comparing the rank with the neighbors rank. (C) The network of 
similarity between software projects cannot be clustered according to either the average 
degree or the network size. The two modules correspond however to networks that were 
built by two independent methods.
}
\label{fig3}
\end{figure}

The $n$-tangle method can also be used to isolate key network features that enable 
classification of networks. In Fig. 3 we present the $n$-tangle connectivity trees 
resulting from a) Facebook friendship networks in 100 Universities in 2005\cite{23}
(described in the Supplemental Materials), b) 
arxiv.org co-authorship\cite{24} in 17 different fields, and c) software code in 14 
different projects\cite{25, 26}. For the Facebook friendship, there is no clear 
clustering with the average degree, but when we consider student enrollment, then we 
discover a similarity between networks at Universities of similar size, at all sizes. The 
$n$-tangle method therefore enables us to obtain meaningful sociological insight into the 
process, where students create online friendships according to the size of the pool of 
possible connections, even though the average number of friendships is much smaller than 
the pool size. It may therefore be that the fundamental nature of the social activities 
and experience is shaped by the total size of the university, even though that number can 
be significantly larger than the size of the average friend-group. For the case of 
co-authorship, on the other hand, the classification of networks according to the network 
size does not work well. We instead discover that the important factor in this case is 
the average degree of an author, i.e. fields with large number of co-authors yield 
similar networks with each other. This classification of networks according to an 
underlying structural property does not trivially result from the $n$-tangle method. In 
the example of software project networks in Fig. 3c we were not able to determine any 
particular structural property that separates the projects in the $n$-tangle networks. 
Interestingly, each of the two modules in Fig. 3c includes software projects that were 
generated by different circumstances. This method therefore, not only allows comparison 
across networks, but enables hypothesis testing about which facets might be the most 
salient organizational features that drive the emergence of networks within the systems 
studied.

\begin{figure}
\centerline{\resizebox{14cm}{!} { \includegraphics{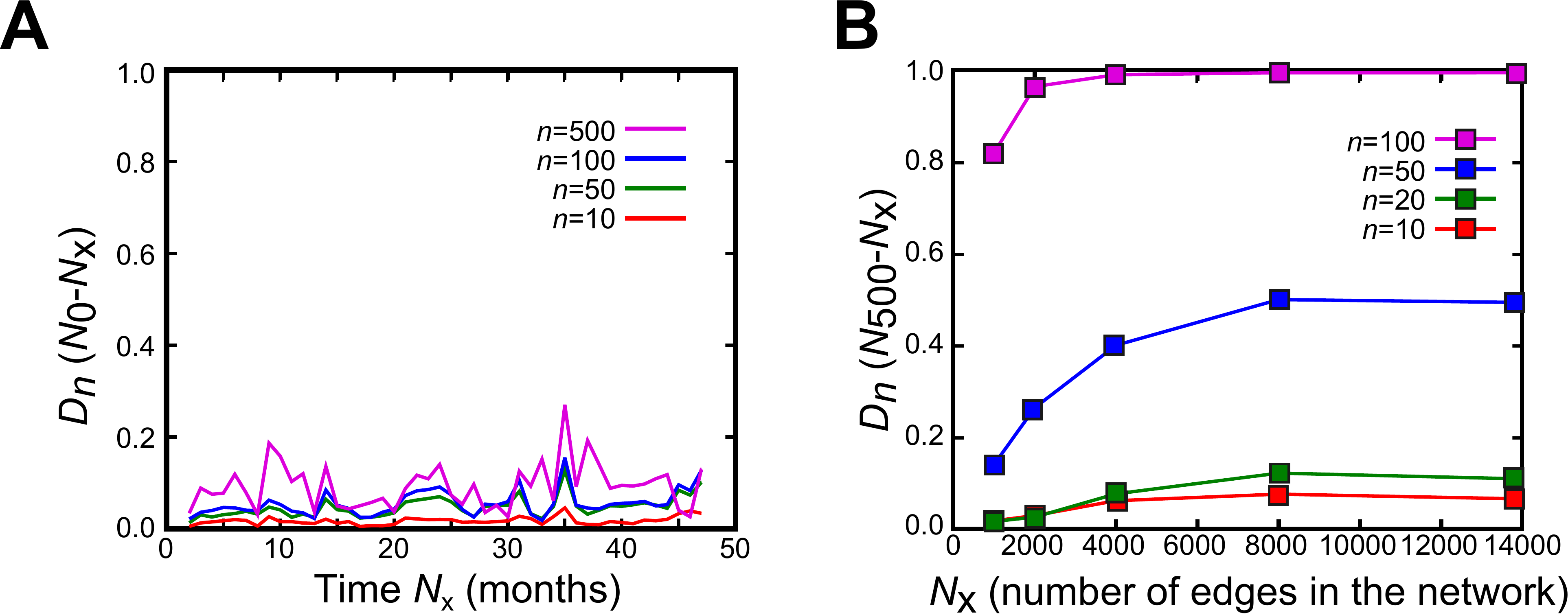}}}
\caption {
Comparison of network evolution. (A) Similarity of the Internet at the AS level with 
itself as a function of time. We compare the KS index $D_n(N_x-N_0)$ of the Internet 
topology at $N_0=$January 2004 with the topology at time $N_x$, which is increasing 
monthly. Independently of the scale $n$, the topology remains the same throughout the 
network evolution for 3 years. (B) Comparison of the KS index in social networking 
friendships as a function of time. We compare the network topology of the early network 
containing 500 links with the networks at subsequent times. The network remains the same 
for small values of $n$, but changes drastically at larger scales.
}
\label{fig4}
\end{figure}

We also applied this method to characterize network evolution. In the examples of the 
Internet growth\cite{27} and online social-networking evolution\cite{28} in Fig. 4, we 
compare the network at a given time with the same network at subsequent times. The 
starting date for the Internet data was January 2004. Our method indicates that the 
Internet topology was already fixed in time by January 2004 and did not change much by 
November 2007, when the network had already doubled in size. This result holds across all 
subgraph sizes, and is also consistent with the macroscopic fact that the average link
density was declining slowly from $2.4$ $10^{-4}$ to $1.4$ $10^{-4}$ over three years.
On the contrary, the facebook-like online network shows a stable behavior 
only at small scales n. The number of edges in the network increases by a factor of 25, 
but the $n$-tangle density remains very similar at any time when $n<20$. When we consider 
larger n values, though, there is a very sharp change between the initial reference 
network and the subsequent instances of it. Therefore, within the same network the 
small-scale structures remain the same, while larger-scale structures evolve into 
different forms. The method can therefore separate structurally stabilized networks over 
time from unstabilized ones. Moreover, in networks of evolving topology we can identify 
differences in the stability of short-scale and larger-scale structures. This may 
therefore enable accurate estimation of the quality of approximation of static snapshots 
of continually shifting networks, which has been shown to be of critical importance in 
areas such as epidemiology\cite{29}.

The calculation of the $n$-tangle density provides a simple and powerful method for 
efficient network comparison.
Understanding the degree of similarity between two 
networks is the key to promote the classification of networks into clusters for further 
analysis of their common features that would otherwise remain unknown, and allows us to 
hypothesize meaningfully about how these clusters may capture fundamental properties of 
networks and the systems they represent.

\begin{acknowledgments}
We thank the Dept. of Homeland Security for funds in support of this research through the 
CCICADA Center at Rutgers, and NSF EaeM grant \#1049088. We are grateful to the following 
people, for making their animal network data available to us: C. Berman, J.R. Madden, J. 
Wolf, D.B. McDonald, P.C. Cross, and S. Sundaresan.
\end{acknowledgments}

\newpage

\pagenumbering{gobble}

\vspace*{-2cm}\hspace*{-3cm}\includegraphics[page=1,width=1.3\textwidth]{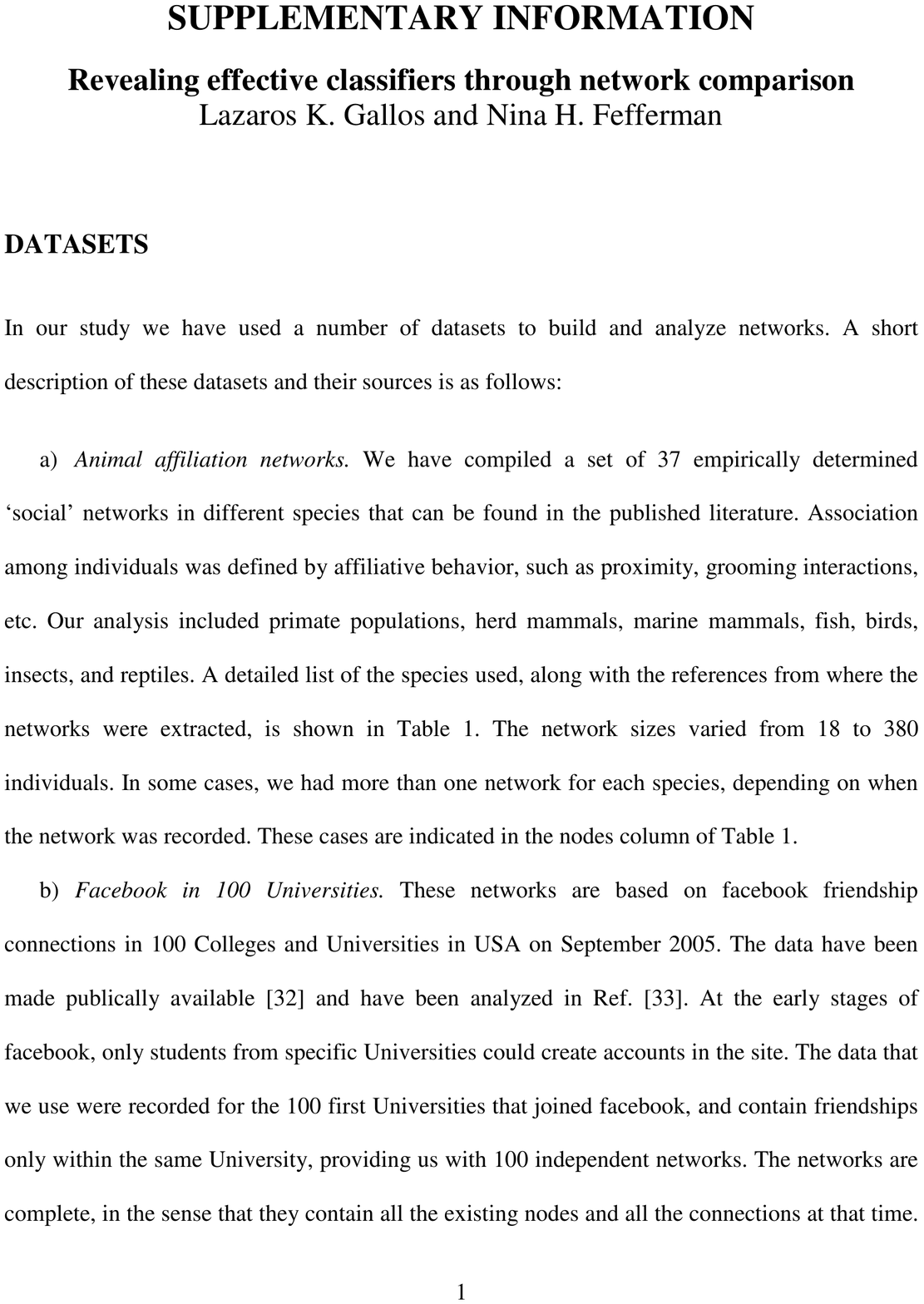}
\clearpage
\vspace*{-2cm}\hspace*{-3cm}\includegraphics[page=2,width=1.3\textwidth]{Supplementary.pdf}
\clearpage
\vspace*{-2cm}\hspace*{-3cm}\includegraphics[page=3,width=1.3\textwidth]{Supplementary.pdf}
\clearpage
\vspace*{-2cm}\hspace*{-3cm}\includegraphics[page=4,width=1.3\textwidth]{Supplementary.pdf}
\clearpage
\vspace*{-2cm}\hspace*{-3cm}\includegraphics[page=5,width=1.3\textwidth]{Supplementary.pdf}
\clearpage
\vspace*{-2cm}\hspace*{-3cm}\includegraphics[page=6,width=1.3\textwidth]{Supplementary.pdf}
\clearpage
\vspace*{-2cm}\hspace*{-3cm}\includegraphics[page=7,width=1.3\textwidth]{Supplementary.pdf}
\clearpage
\vspace*{-2cm}\hspace*{-3cm}\includegraphics[page=8,width=1.3\textwidth]{Supplementary.pdf}
\clearpage
\vspace*{-2cm}\hspace*{-3cm}\includegraphics[page=9,width=1.3\textwidth]{Supplementary.pdf}

\end{document}